\documentclass[twocolumn,showpacs,showkeys,groupedaddress]{revtex4-1}

\usepackage{graphicx}
\usepackage{amsmath,amsthm,amssymb}
\usepackage[toc,page]{appendix}
\usepackage[T1]{fontenc}
\usepackage{siunitx}
\usepackage{microtype}
\usepackage{hyperref}
\usepackage{braket}
\usepackage{color,soul}

\usepackage{fancyhdr} % to change header and footers
\fancypagestyle{plain}{%
    \fancyhf{}%
    \fancyfoot[R]{\thepage}%
}

\renewcommand\appendix{\par
  \setcounter{section}{0}
  \setcounter{subsection}{0}
  \setcounter{figure}{0}
  \setcounter{table}{0}
  \renewcommand\thesection{Appendix \Alph{section}}
  \renewcommand\thefigure{\Alph{section}\arabic{figure}}
  \renewcommand\thetable{\Alph{section}\arabic{table}}
}
\newcommand*{\hatH}{\hat{\mathcal{H}}}

\graphicspath{{./figures/}}

\begin{document}

\title{Quantum interference of topological states of light}

\author{Jean-Luc Tambasco$^1$, Giacomo Corrielli$^{2,3}$, Robert J. Chapman$^1$, Andrea Crespi$^{2,3}$, Oded Zilberberg$^4$, Roberto Osellame$^{2,3}$, Alberto Peruzzo$^1$}
\affiliation{
$^1$Quantum Photonics Laboratory and Centre for Quantum Computation and Communication Technology, School of Engineering, RMIT University, Melbourne, Victoria 3000, Australia \\
$^2$Istituto di Fotonica e Nanotecnologie, Consiglio Nazionale delle Ricerche, Piazza Leonardo da Vinci 32, Milano I-20133, Italy \\
$^3$Dipartimento di Fisica, Politecnico di Milano, Piazza Leonardo da Vinci 32, Milano I-20133, Italy \\
$^4$Institute for Theoretical Physics, ETH Zurich, 8093 Zurich, Switzerland
}
\date{\today}

\begin{abstract}
\noindent
Topological insulators are materials that have a gapped bulk energy spectrum, but contain protected in-gap states appearing at their surface. These states exhibit remarkable properties such as unidirectional propagation and robustness to noise that offer an opportunity to improve the performance and scalability of quantum technologies. For quantum applications, it is essential that the topological states are indistinguishable.
Here we report high-visibility quantum interference of single photon topological states in an integrated photonic circuit. 
Two topological boundary-states, initially at opposite edges of a coupled waveguide array, are brought into proximity, where they interfere and undergo a beamsplitter operation.
We observe $93.1\pm2.8\%$ visibility Hong-Ou-Mandel (HOM) interference, a hallmark non-classical effect that is at the heart of linear optics-based quantum computation.
Our work shows that it is feasible to generate and control highly indistinguishable single photon topological states, opening pathways to enhanced photonic quantum technology with topological properties, and to study quantum effects in topological materials.
\end{abstract}

%\pacs{}
%\keywords{topological insulators, quantum interference, integrated photonics}

\maketitle

\setcounter{secnumdepth}{1}

\section*{Introduction}

\begin{figure*}[!t]
\centering
\includegraphics[width=1.0\linewidth]{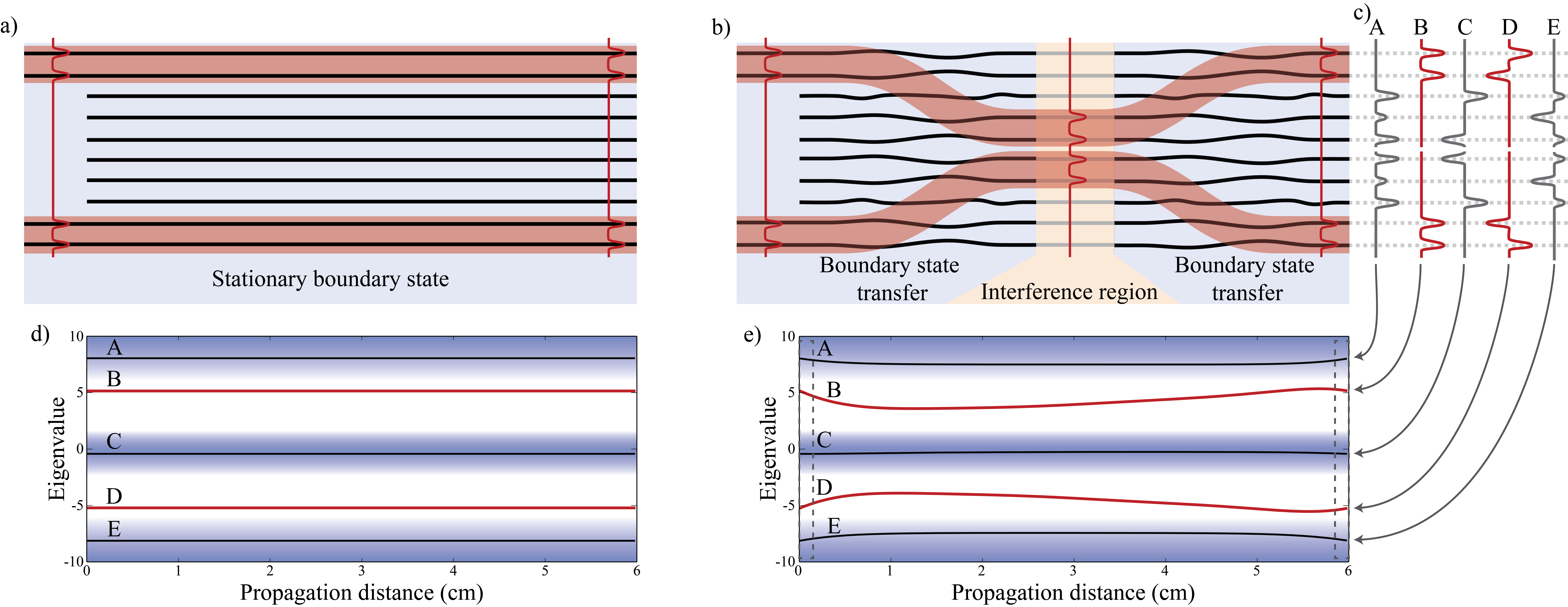}
\caption{
\textbf{Photonic boundary-state beamsplitter.}
(a) Illustrative representation of a waveguide array implementing stationary topologically boundary-states (red shaded regions) that propagate at the edges of the device. This device is used to confirm that the boundary-state is preserved during the propagation inside the array.
(b) Illustrative representation of a waveguide array implementing a `topological beamsplitter' that interferes two topologically boundary-states.
(c) Photonic supermodes (eigenvectors) of the arrays at the start and end of the both devices.
(d, e) Band structure (eigenenergies) along the length of the arrays a) and b).
The topological bands (B and D) are highlighted in red and the bulk bands (A, C and E) are shaded in blue.
}
\label{fig:device}
\end{figure*}

Research into solid-state physics has led to the discovery of a new phase of matter, the \emph{topological insulator}; a class of materials that insulates in the bulk, but conducts on the surface \cite{hasan_colloquium:_2010,RevModPhys.83.1057}. This has inspired the design of new topological systems with unique band structures and protected boundary-states in various effective dimensions. In particular, one-dimensional topological superconductors have recently received great attention due to their topological boundary state, namely Majorana zero-modes that can be harnessed for topological quantum computing \cite{lutchyn_realizing_2017}.

Since the discovery of topological phases of matter, a wealth of pioneering topological systems have been demonstrated using photonics~\cite{lu_topological_2014,ozawa2018topological}.
Topological photonics has the advantage of not requiring strong magnetic fields, and features intrinsically high-coherence, room-temperature operation and easy manipulation. 
To-date, several topological effects have been observed using integrated photonics including Majorana modes \cite{Xu2016}, chiral edge modes robust to defects \cite{wang_observation_2009, hafezi_robust_2011, rechtsman_photonic_2013, hafezi_imaging_2013, blanco-redondo_topological_2016, mittal_measurement_2016, xiao_observation_2017}, optical Weyl points \cite{Chen2016, Noh:2017be, Li:2017cx}, 1D and 2D topological pumping and topological quasicrystals \cite{kraus_topological_2012, PhysRevB.91.064201, PhysRevLett.110.076403, zilberberg_photonic_2017}, as well as generation and propagation of single photons \cite{blanco-redondo_photonic_2017, mittal_topologically_2017}.

Concurrently, photonics has a long-standing goal to implement quantum computation.
Quantum interference of single photons at a 50:50 beamsplitter is a key phenomenon in quantum physics and lies at the heart of linear-optical quantum computation \cite{Obrien:2009eu}. This phenomenon can be observed via the well-known Hong-Ou-Mandel (HOM) experiment \cite{hong_measurement_1987}, which has been demonstrated in integrated photonic devices, including on-chip beamsplitters \cite{politi_silica--silicon_2008, laing_high-fidelity_2010, peruzzo_multimode_2011}, photonic quantum walks \cite{peruzzo_quantum_2010, sansoni_two-particle_2012}, circuits displaying Anderson localization \cite{crespi_anderson_2013}, and recently in plasmonic devices \cite{heeres_quantum_2013}. 
High-visibility quantum interference relies on the two input photons being totally indistinguishable.
To-date, quantum interference has not been observed in topological systems.

In this work, we report high-visibility quantum interference of two single-photon topological boundary-states in a photonic waveguide array.
We engineered a time-varying Hamiltonian, controlling the band structure of the device and the spatial isolation of the topological states to implement a 50:50 beamsplitter.
Using this `topological beamsplitter', we measured Hong-Ou-Mandel interference with $93.1\pm2.8\%$ visibility, demonstrating non-classical behavior of topological states.

\section*{Results}

Our device implements the off-diagonal Harper model, which describes a one-dimensional lattice that exhibits topological boundary-states \cite{harper_single_1955, kraus_topological_2012, kraus_topological_2012_a}.
The time-varying Hamiltonian of this model is given as
\begin{equation}
\hatH(t) = \sum_{n=1}^{N-1} \kappa_n(t) (\hat{a}_n \hat{a}_{n+1}^\dag + \hat{a}_n^\dag \hat{a}_{n+1}),
\label{eq:Hamiltonian}
\end{equation}
where $\hat{a}_n$ and $\hat{a}_{n}^\dag$ are annihilation and creation operators acting on lattice site $n$. $\kappa_n(t)$ is the coupling strength at time $t$ between site $n$ and site $n + 1$.
In the off-diagonal Harper model, the coupling strengths follow the periodic function
\begin{equation}
\kappa_n(t) = \kappa_0 \left[1 + \Lambda(t)\cos\left(2 \pi \bar{b} n + \phi(t)\right) \right],
\label{eq:Harper}
\end{equation}
where $\kappa_0$ is the nominal coupling coefficient between two adjacent lattice sites, $\bar{b}$ controls the periodicity of the lattice, $\Lambda(t)$ controls the size of the spectral gaps and correspondingly defines the confinement of the boundary-state at time $t$, and $\phi(t)$ is a time-varying phase.
By carefully choosing the value of $\bar{b}$, gaps appear in the energy spectrum of the system that allow topological pumping by adiabatically varying $\phi(t)$~\cite{kraus_topological_2012}. 
Carefully choosing $\phi(t)$ ensures the appearance of topological boundary modes on the edges of the array.
Both $\phi(t)$ \cite{kraus_topological_2012} and $\Lambda(t)$ can be used to manipulate the boundary-states.
In this work we vary $\Lambda(t)$ to confine, delocalize and interfere the boundary-states; this procedure is reminiscent to changing the length of a topological superconductor and interfering its Majorana-modes~\cite{lutchyn_realizing_2017}.

An array of coupled waveguides in the nearest neighbor approximation implements the same tight-binding Hamiltonian as Eq. \eqref{eq:Hamiltonian}, where the waveguide separation controls the coupling strength.
We experimentally characterized the relationship between the waveguide separation and the coupling strength $\kappa_n(t)$ (see Materials and Methods for details), which enabled us to design an array with the desired Hamiltonian.
Because we vary the $\kappa_n$ terms along the \emph{length} of the array, we make the transformation from a time-varying to a distance-varying Hamiltonian with the relationship $z = \tfrac{ct}{n}$ where $z$ is the position, $c$ is the speed of light and $n$ is the waveguide effective refractive index.

\begin{figure*}[t!]
\centering
\includegraphics[width=0.9\textwidth]{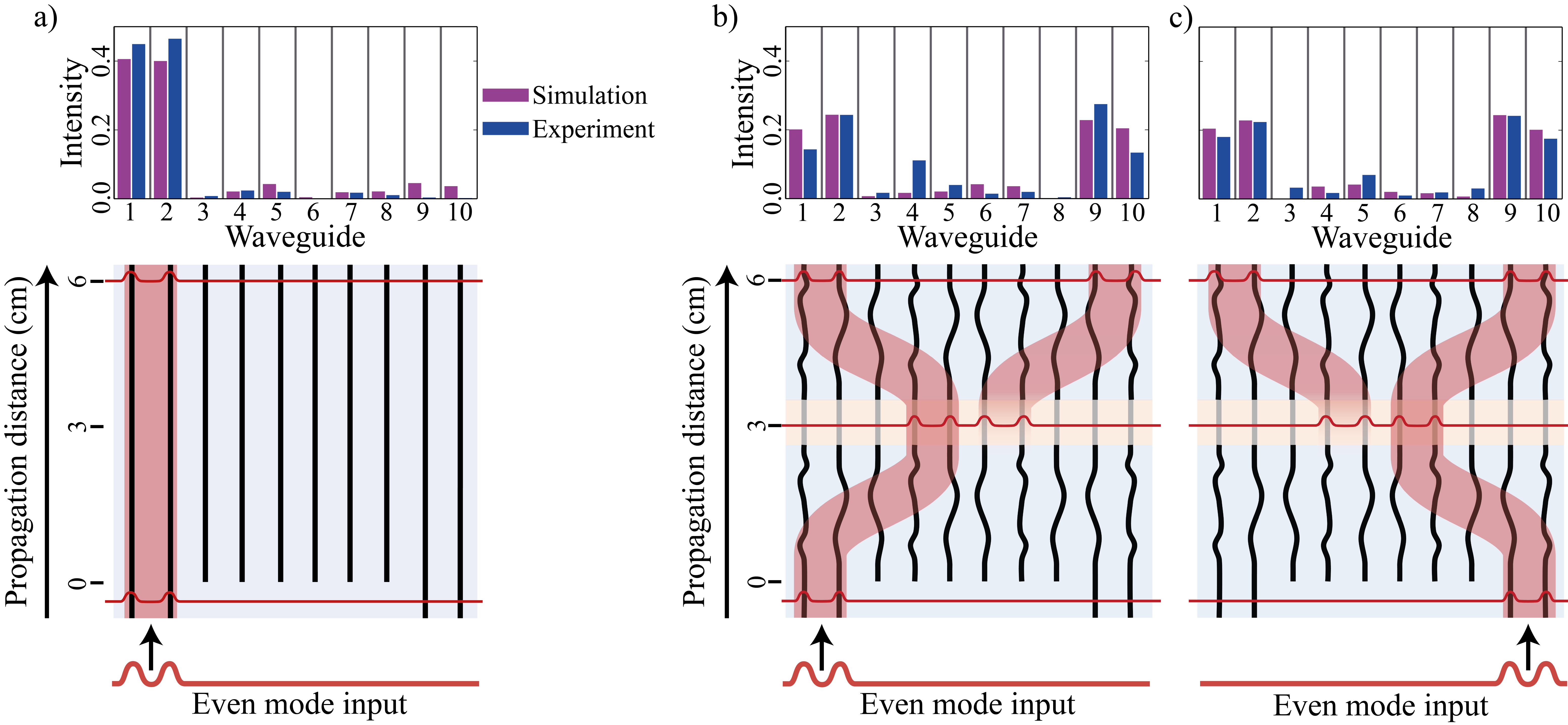}
\caption{
\textbf{Characterization of the stationary boundary-state device and the topological beamsplitter.}
The output of the chip is characterized using laser light and a CCD camera.
(a) The normalized output intensity distribution of the stationary boundary-state.
(b, c) The normalized output intensity distribution of the topological beamsplitter with injection into the left and right inputs respectively.
}
\label{fig:classical_results}
\end{figure*}

Initially, two photonic states are localized at the edges of the array; they are spectrally and spatially isolated from the bulk modes, start with the same energy, and are spatially isolated from one another. 
Interference can occur when these states are adiabatically delocalized from the edges to the bulk of the lattice by reducing the bulk gap size.

We designed two devices, each consisting of ten waveguides with symmetric coupling strengths $\{\kappa_1, \kappa_2, \kappa_3, \kappa_4, c, \kappa_4, \kappa_3, \kappa_2, \kappa_1\}$ and $\bar{b} = 2/3$.
The first device has fixed $\Lambda(z)=0.6$ to demonstrate and confirm the confinement of the topological boundary-states, as illustrated in Fig. \ref{fig:device}(a).
In the second device, illustrated in Fig. \ref{fig:device}(b), we vary $\Lambda(z)$ from $\Lambda(0)=0.6$ to $\Lambda(L/2)=0.1$, where $L$ is the total length of the array.
This reduces the localization of the two boundary modes, causing them to interfere.
By tuning the central coupling coefficient ($c$), a 50:50 beamsplitter is realized before relocalizing the states to the sides of the device, where $\lambda(L)=0.6$.
For both devices, waveguides 1--5 (and due to symmetry, 10--6) have five photonic supermodes (eigenstates), which are shown in Fig. \ref{fig:device}(c).

Exciting the boundary-states of each array requires injecting into the mode labeled B in Fig. \ref{fig:device}(c).
As shown in Figs. \ref{fig:device}(a) and (b), this is achieved by extending the two edge waveguides to the input facet of the chip (see Supplementary Section S1 for details).
To model the bulk-band spectrum of the photonic supermodes in Fig. \ref{fig:device}(c), we calculated the eigenvalues of both devices, shown in Figs. \ref{fig:device}(d) and (e), along the length of the array.
The approximate bulk energy bands are shaded in blue and the eigenvalues corresponding to the boundary-states are plotted in red (labeled B and D).
As the Hamiltonian is implemented on a photonic platform, each eigenvalue is proportional to the effective refractive index of the corresponding photonic supermode \cite{Huang:94}.
If eigenvalues are similar in magnitude, the corresponding eigenstates will scatter between modes; however, eigenstates between the energy bands are resilient to scattering.

Our devices were fabricated using the direct-write technique \cite{gattass_femtosecond_2008, valle_micromachining_2009} as it enables high-precision control of the waveguide coupling coefficients.
The direct-write technique is implemented by translating a borosilicate chip while focusing a femtosecond laser into the bulk (see Methods and Materials for more details on the chip fabrication).

We characterized each device using laser light at \SI{808}{\nano\metre}, to match the wavelength of our single photon source, and measured the output with a CCD camera.
We calculated the fidelity between the measured output distribution across the whole array and the simulated result as $F=\sum_i\sqrt{P_i^{\mathrm{S}}P_i^{\mathrm{M}}}$, where $P_i^{\mathrm{S}}$ ($P_i^{\mathrm{M}}$) is the simulated (measured) intensity of light at the output of waveguide $i$ after normalization.
The intensity distribution is equivalent to the output single-photon probability distribution.
Here the simulation is based on the physical parameters of our device.
We note that depicting the boundary-state supermodes (B and D in Fig \ref{fig:device}(c)) as being confined to two-waveguide is an approximation and, in a real device, they exponentially decay beyond the edge waveguides---this phenomenon is inherent to any bound mode in a spectral gap.

Figure \ref{fig:classical_results}(a) shows the measured output intensity and simulation results for the stationary topological boundary-state when injecting into the left even-mode eigenstate, and the fidelity is $F=97.1\%$.
Figure \ref{fig:classical_results}(b) shows the results for the topological beamsplitter.
We measured fidelity for the left and right input of $F=96.3\%$ and $F=97.8\%$ respectively.
These fidelities are very high and are mainly limited by fabrication imperfections.

\begin{figure*}[!t]
\centering
\includegraphics[width=1.0\linewidth]{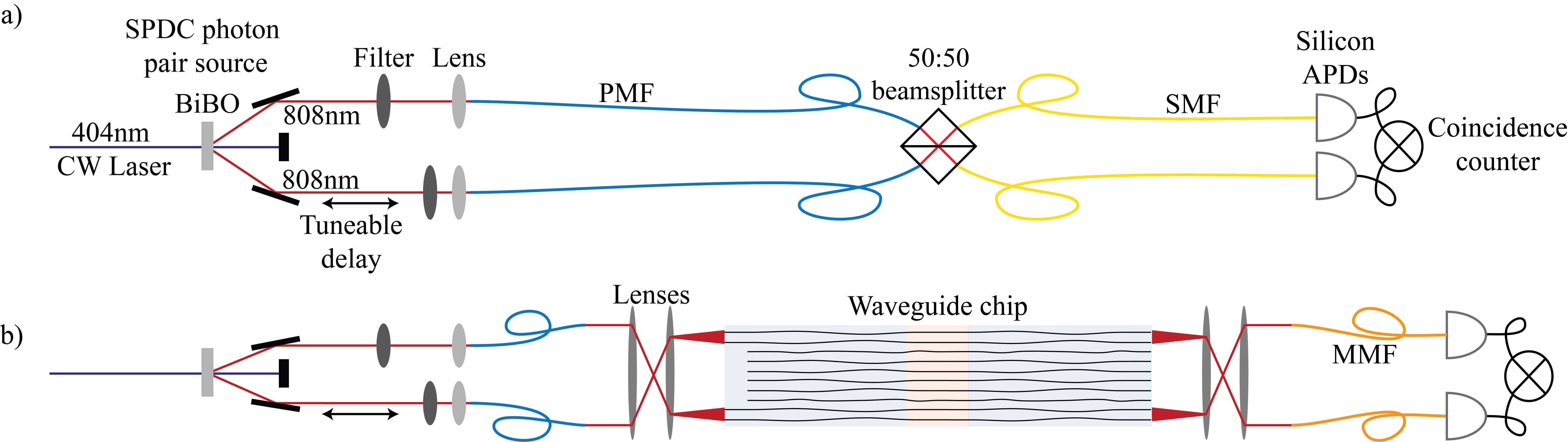}
\caption{
\textbf{Experimental setup for interfering topological boundary-states.}
(a) Setup to characterize the indistinguishably of the photon pairs generated from a spontaneous parametric down-conversion (SPDC) source.
The photons are interfered in a 50:50 beamsplitter via polarization maintaining  fibers (PMF).
The output of the beamsplitter is pigtailed with single mode fiber (SMF) connected to single photon avalanche photodiodes (APDs).
Coincidence counts are measured between the two detectors with a timing card.
(b) To perform the indistinguishability measurements of single photon topologically protected states, the pigtailed beamsplitter in a) is replaced with the topological beamsplitter device.
We used PMF, multimode fibers (MMF) and free-space lenses to couple photons to the device.
}
\label{fig:fig_3}
\end{figure*}

\begin{figure}[!t]
\centering
\includegraphics[width=0.9\linewidth]{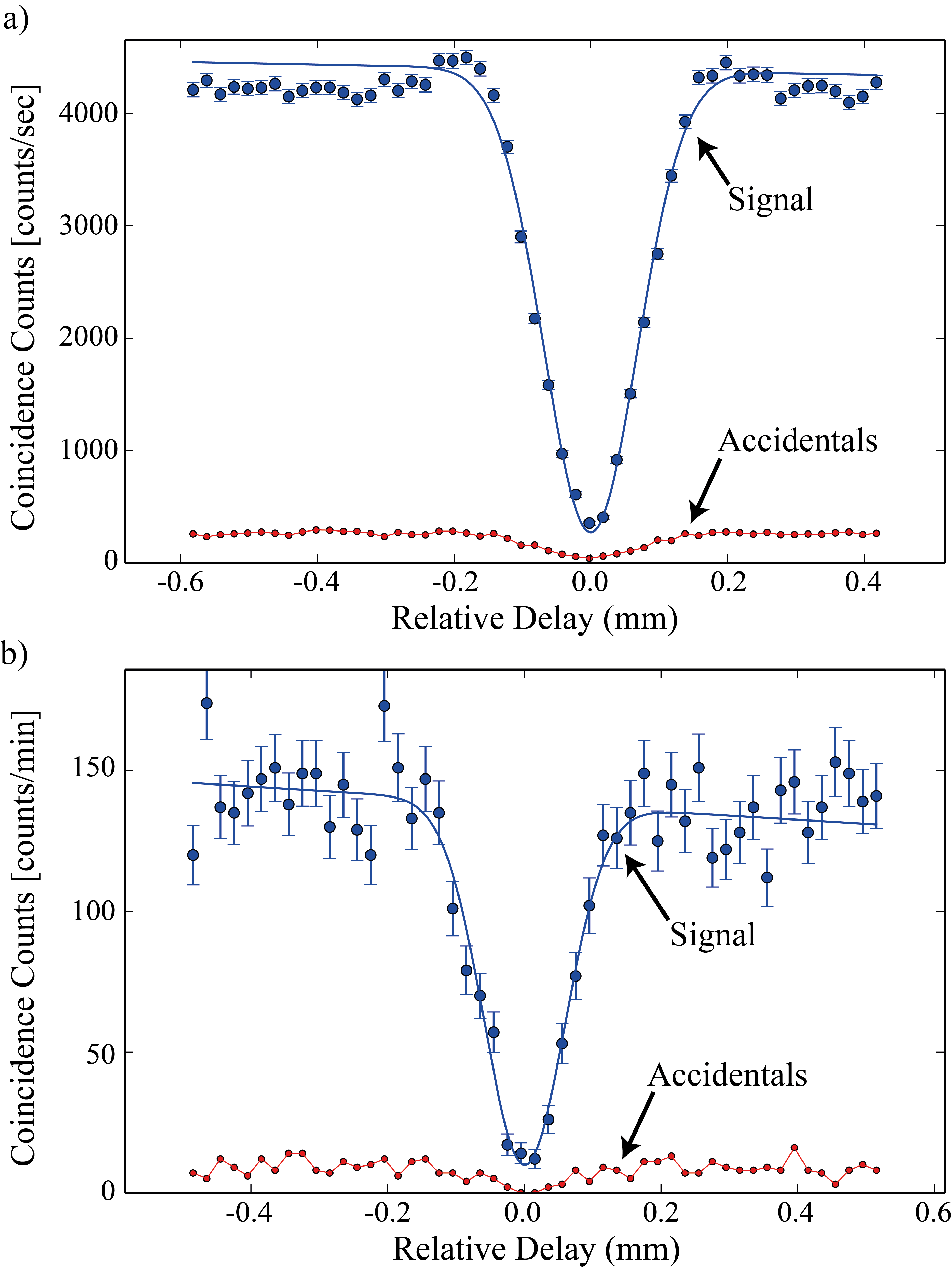}
\caption{\textbf{Measurements of indistinguishability.} a) HOM interference of single photons using a commercially available fiber pigtailed 50:50 beamsplitter with a visibility of $94.5\pm0.5\%$. b) HOM interference on the topological beamsplitter with a visibility of $93.1\pm2.8\%$. The error bars shown are based on Poissonian statistics.}
\label{fig:hom}
\end{figure}

In the quantum interference experiment, we employed a coupling setup to collect photons from the outer waveguides (1,2 and 9,10).
When selecting only these waveguides, we calculated the reflectivity of the topological beamsplitter to be 49.9\% (50.7\%) for the left (right) input, which is very close to 50\%---a requirement for high visibility quantum interference.

Figure \ref{fig:fig_3}(a) shows our setup for measuring HOM interference.
We generated pairs of photons using a free-space spontaneous parametric down conversion (SPDC) source before coupling into two polarization maintaining optical fibers (PMF).
Narrow-band filters are inserted to ensure the photon wavelengths are matched and one fiber is positioned on a translation stage to enable a tunable delay (see Supplementary Section S2 for full details on the photon pair source).
We measure the visibility of the interference by controlling the distinguishability of the photons with the tunable delay.

We first injected single photon pairs into a commercially available fiber coupled 50:50 beamsplitter (FBS) with PMF for the input, ensuring the photons have the same polarization when they interfere.
The output fibers are coupled to single photon avalanche photodiodes (APDs) that emit an electrical pulse when a photon is detected.
Coincidence measurements of the APD signals are performed with a timing card.
We measured the HOM dip shown in Fig \ref{fig:hom}(a) with visibility $V_{\mathrm{FBS}}=94.5\pm0.5\%$.
Error bars on the plot are calculated using Poissonian statistics (see Supplementary Section S3 for HOM dip error calculation).
Accidental coincidences due to stray ambient light and dark counts were detected and subtracted from the signal by applying an electronic time delay to one detector.
As the beamsplitter reflectivity is close to ideal ($r=49.0\pm0.1\%$), the visibility is limited predominately by the spectral distinguishability of the generated photon pairs.

We then replaced the 50:50 fiber beamsplitter with our waveguide chip, as shown in Fig. \ref{fig:fig_3}(b).
We injected single photons simultaneously into both boundary-states of the topological beamsplitter (TBS) and varied the delay such that we could perfectly match the arrival times.
We measured the HOM dip shown in Fig. \ref{fig:hom}(b) with a visibility of $V_{\mathrm{TBS}}=93.1\pm2.8\%$; this gives a relative visibility $V_\mathrm{relative}=\tfrac{V_{\mathrm{TBS}}}{V_{\mathrm{FBS}}}$ of $98.5\pm3.5\%$, confirming that the quantum interference of topological boundary-states in our device is close to ideal.
The measurement noise for the topological beamsplitter chip is increased due to coupling losses leading to a significantly lower count rate and, consequently, a decreased signal-to-noise ratio.

\section*{Discussion} 
In this work, we have demonstrated that single photons localized to topological boundary-states can undergo high-visibility quantum interference.
To this aim, we employed a laser-written photonic circuit that represents one of the most complex examples of a continuous waveguide array with engineered coupling coefficients varying along the propagation direction.
This technology enables future studies of quantum effects in topological materials that are challenging or impossible to probe due to, for example, large magnetic field requirements or excessive noise \cite{RevModPhys.83.1057}.  Moreover, the TBS could be extended to other topological models (such as the Su, Schrieffer and Heeger (SSH) model of a one-dimensional dimer chain \cite{su_solitons_1979}).
We anticipate that the TBS presented in this work will combine with other leading works in topological photonics \cite{mittal_topologically_2017,hafezi_robust_2011} to help solve challenges currently faced in quantum photonics, including pump filtering for photon generation and robust photon transport.

\noindent{\bf Acknowledgments: }
G.C., A.C. and R.O. acknowledge financial support by the ERC-Advanced Grant CAPABLE (Composite integrated photonic platform by femtosecond laser micromachining; grant agreement no. 742745) and by the H2020-FETPROACT-2014 Grant QUCHIP (Quantum Simulation on a Photonic Chip; grant agreement no. 641039). 
O.Z. acknowledges financial support from the Swiss National Science Foundation (SNSF). 
A.P. acknowledges support from the Australian Research Council Centre of Excellent for Quantum Computation and Communication Technology (CQC$^2$T), project no. CE170100012, an Australian Research Council Discovery Early Career Researcher Award, project no. DE140101700 and an RMIT University Vice-Chancellor`s Senior Research Fellowship.

\section*{Materials And Methods}
\subsection{Device design and simulation}

The relationship between waveguide separation and coupling coefficient is characterized with a test chip containing varying spaced waveguides.
This relationship follows an exponential decay $\kappa = ae^{-bd}$, where $a = 115$ cm$^{-1}$ and $b = 0.36$ $\mu$m$^{-1}$ are experimentally measured constants and $d$ is the separation between the waveguides.
We can invert this function to find the waveguide separation necessary to achieve the desired coupling coefficients in Eq. \ref{eq:Harper}.
These $\kappa_n(z)$ coupling coefficients control the transfer of the topological boundary-state from the sides of the array to the center.

We numerically optimize the coupling coefficient $c$ in Eq. \ref{eq:Hamiltonian} such that the boundary-states couple with 50\% probability.
This implements a 50:50 beamsplitter operation.

Finally, the waveguide separations are adjusted to transfer the boundary-states back to the sides of the array.

\subsection{Device fabrication}
We employ the femtosecond direct-write technique for fabricating waveguides in borosilicate glass \cite{gattass_femtosecond_2008,valle_micromachining_2009}.
Our SPDC source generates photons close to 808 nm wavelength and we fabricate waveguides that are single mode at this wavelength.
The waveguides are fabricated by focusing a femtosecond pulsed laser with a repetition rate of 1 $\mathrm{MHz}$ and energy of 220 nJ/pulse in the bulk of a borosilicate substrate (Eagle2000, Corning) by means of a 50$\times$ microscope objective (NA = 0.6).
Waveguides are patterned by translating the sample at the constant speed of 40 $\mathrm{mm/s}$.

The resulting waveguides exhibit relatively low propagation losses (0.5 $\mathrm{dB/cm}$) and a slightly elliptical guided mode, with an average diameter of $\sim$8$\mathrm{\mu m}$.

The separation between neighboring waveguides controls the rate of coupling.
There is, inevitably, coupling between next-nearest-neighbor waveguides, however, the coupling decays exponentially with distance and as such, we can approximate to a nearest-neighbor model.


\begin{thebibliography}{10}

\bibitem{hasan_colloquium:_2010}
M.~Z. Hasan, C.~L. Kane, {\it Reviews of Modern Physics\/} {\bf 82}, 3045
  (2010).

\bibitem{RevModPhys.83.1057}
X.-L. Qi, S.-C. Zhang, {\it Rev. Mod. Phys.\/} {\bf 83}, 1057 (2011).

\bibitem{lutchyn_realizing_2017}
R.~M. Lutchyn, {\it et~al.\/}, {\it arXiv:1707.04899\/}  (2017).

\bibitem{lu_topological_2014}
L.~Lu, J.~D. Joannopoulos, M.~Solja{\v c}i{\'c}, {\it Nature Photonics\/} {\bf
  8}, 821 (2014).

\bibitem{ozawa2018topological}
T.~Ozawa, {\it et~al.\/}, {\it arXiv preprint arXiv:1802.04173\/}  (2018).

\bibitem{Xu2016}
J.-S. Xu, {\it et~al.\/}, {\it Nat Commun\/} {\bf 7}, 13194 (2016).

\bibitem{wang_observation_2009}
Z.~Wang, Y.~Chong, J.~D. Joannopoulos, M.~Solja{\v c}i{\'c}, {\it Nature\/}
  {\bf 461}, 772 (2009).

\bibitem{hafezi_robust_2011}
M.~Hafezi, E.~A. Demler, M.~D. Lukin, J.~M. Taylor, {\it Nature Physics\/} {\bf
  7}, 907 (2011).

\bibitem{rechtsman_photonic_2013}
M.~C. Rechtsman, {\it et~al.\/}, {\it Nature\/} {\bf 496}, 196 (2013).

\bibitem{hafezi_imaging_2013}
M.~Hafezi, S.~Mittal, J.~Fan, A.~Migdall, J.~M. Taylor, {\it Nature
  Photonics\/} {\bf 7}, 1001 (2013).

\bibitem{blanco-redondo_topological_2016}
A.~Blanco-Redondo, {\it et~al.\/}, {\it Physical Review Letters\/} {\bf 116},
  163901 (2016).

\bibitem{mittal_measurement_2016}
S.~Mittal, S.~Ganeshan, J.~Fan, A.~Vaezi, M.~Hafezi, {\it Nature Photonics\/}
  {\bf 10}, 180 (2016).

\bibitem{xiao_observation_2017}
L.~Xiao, {\it et~al.\/}, {\it Nature Physics\/} {\bf 13}, 1117 (2017).

\bibitem{Chen2016}
W.-J. Chen, M.~Xiao, C.~T. Chan, {\it Nature Communications\/} {\bf 7}, 13038
  (2016).

\bibitem{Noh:2017be}
J.~Noh, {\it et~al.\/}, {\it Nature Physics\/} {\bf 349}, 622 (2017).

\bibitem{Li:2017cx}
F.~Li, X.~Huang, J.~Lu, J.~Ma, Z.~Liu, {\it Nature Physics\/} {\bf 5}, 031013
  (2017).

\bibitem{kraus_topological_2012}
Y.~E. Kraus, Y.~Lahini, Z.~Ringel, M.~Verbin, O.~Zilberberg, {\it Physical
  Review Letters\/} {\bf 109}, 106402 (2012).

\bibitem{PhysRevB.91.064201}
M.~Verbin, O.~Zilberberg, Y.~Lahini, Y.~E. Kraus, Y.~Silberberg, {\it Phys.
  Rev. B\/} {\bf 91}, 064201 (2015).

\bibitem{PhysRevLett.110.076403}
M.~Verbin, O.~Zilberberg, Y.~E. Kraus, Y.~Lahini, Y.~Silberberg, {\it Phys.
  Rev. Lett.\/} {\bf 110}, 076403 (2013).

\bibitem{zilberberg_photonic_2017}
O.~Zilberberg, {\it et~al.\/}, {\it arXiv:1705.08361\/}  (2017).

\bibitem{blanco-redondo_photonic_2017}
A.~Blanco-Redondo, B.~Bell, M.~Segev, B.~J. Eggleton, {\it AIP Conference
  Proceedings\/} {\bf 1874}, 020001 (2017).

\bibitem{mittal_topologically_2017}
S.~Mittal, M.~Hafezi, {\it arXiv:1709.09984\/}  (2017).

\bibitem{Obrien:2009eu}
J.~L. O'Brien, A.~Furusawa, J.~Vuckovic, {\it Nature Photonics\/} {\bf 3}, 687
  (2009).

\bibitem{hong_measurement_1987}
C.~K. Hong, Z.~Y. Ou, L.~Mandel, {\it Physical Review Letters\/} {\bf 59}, 2044
  (1987).

\bibitem{politi_silica--silicon_2008}
A.~Politi, M.~J. Cryan, J.~G. Rarity, S.~Yu, J.~L. O'Brien, {\it Science\/}
  {\bf 320}, 646 (2008).

\bibitem{laing_high-fidelity_2010}
A.~Laing, {\it et~al.\/}, {\it Applied Physics Letters\/} {\bf 97}, 211109
  (2010).

\bibitem{peruzzo_multimode_2011}
A.~Peruzzo, A.~Laing, A.~Politi, T.~Rudolph, J.~L. O'Brien, {\it Nature
  Communications\/} {\bf 2}, 224 (2011).

\bibitem{peruzzo_quantum_2010}
A.~Peruzzo, {\it et~al.\/}, {\it Science\/} {\bf 329}, 1500 (2010).

\bibitem{sansoni_two-particle_2012}
L.~Sansoni, {\it et~al.\/}, {\it Physical Review Letters\/} {\bf 108}, 010502
  (2012).

\bibitem{crespi_anderson_2013}
A.~Crespi, {\it et~al.\/}, {\it Nature Photonics\/} {\bf 7}, 322 (2013).

\bibitem{heeres_quantum_2013}
R.~W. Heeres, L.~P. Kouwenhoven, V.~Zwiller, {\it Nature Nanotechnology\/} {\bf
  8}, 719 (2013).

\bibitem{harper_single_1955}
P.~G. Harper, {\it Proceedings of the Physical Society. Section A\/} {\bf 68},
  874 (1955).

\bibitem{kraus_topological_2012_a}
Y.~E. Kraus, O.~Zilberberg, {\it Physical Review Letters\/} {\bf 109}, 116404
  (2012).

\bibitem{Huang:94}
W.-P. Huang, {\it Journal of the Optical Society of America A\/} {\bf 11}, 963
  (1994).

\bibitem{gattass_femtosecond_2008}
R.~R. Gattass, E.~Mazur, {\it Nature Photonics\/} {\bf 2}, 219 (2008).

\bibitem{valle_micromachining_2009}
G.~D. Valle, R.~Osellame, P.~Laporta, {\it Journal of Optics A: Pure and
  Applied Optics\/} {\bf 11}, 013001 (2009).

\bibitem{su_solitons_1979}
W.~P. Su, J.~R. Schrieffer, A.~J. Heeger, {\it Physical Review Letters\/} {\bf
  42}, 1698 (1979).

\end{thebibliography}
\end{document}